\documentclass[prd,twocolumn,amsmath,amssymb,nofootinbib,floatfix,superscriptaddress]{revtex4}

\usepackage{graphicx,bm,float}

\makeatletter
\def\graphicscale{\twocolumn@sw{0.3}{0.4}}
\def\graphicthreescale{\twocolumn@sw{0.3}{0.4}}

\begin{document}

\title{Two-dimensional 
multicomponent  Abelian-Higgs lattice models}

\author{Claudio Bonati} 
\affiliation{Dipartimento di Fisica dell'Universit\`a di Pisa 
       and INFN Largo Pontecorvo 3, I-56127 Pisa, Italy}

\author{Andrea Pelissetto}
\affiliation{Dipartimento di Fisica dell'Universit\`a di Roma Sapienza
        and INFN Sezione di Roma I, I-00185 Roma, Italy}

\author{Ettore Vicari} 
\affiliation{Dipartimento di Fisica dell'Universit\`a di Pisa
       and INFN Largo Pontecorvo 3, I-56127 Pisa, Italy}

\date{\today}

\begin{abstract}
We study the two-dimensional lattice multicomponent Abelian-Higgs model,
which is a lattice compact U(1) gauge theory coupled with an 
$N$-component complex scalar
field, characterized by a global SU($N$) symmetry.  
In agreement with the Mermin-Wagner theorem, the model has only
a disordered phase at finite temperature and a critical behavior is 
only observed in the zero-temperature
limit.  The universal features are investigated by numerical analyses of the
finite-size scaling behavior in the zero-temperature limit.  The
results show that the renormalization-group flow of the 2D lattice
$N$-component Abelian-Higgs model is asymptotically controlled by the
infinite gauge-coupling fixed point, associated with the universality
class of the 2D CP$^{N-1}$ field theory.
\end{abstract}

\maketitle

% ========================= BODY =========================

\section{Introduction}
\label{intro}

Models of complex scalar matter fields with abelian and nonabelian
gauge symmetries effectively emerge in several interesting systems,
such as superconductors and superfluids, quantum Hall states, quantum
SU($N$) antiferromagnets, unconventional quantum phase transitions,
etc., see, {\em e.g.},
Refs.~\cite{HLM-74,SSNHS-03,SSS-04,BSA-04,MV-04,SBSVF-04,KHI-11,%%
  WNMXS-17,GASVW-18,SSST-19,GSF-19,Sachdev-19} and references therein.
Among the paradigmatic models considered, an important role is played
by the multicomponent lattice Abelian-Higgs (AH) model or lattice
scalar electrodynamics, which is a lattice U(1) gauge theory coupled with an
$N$-component complex scalar field, characterized by a
global SU($N$) symmetry. Its phase diagram has been
thoroughly investigated in three dimensions, considering both the
compact and noncompact formulation of electrodynamics, 
see,
e.g., Refs.~\cite{SSNHS-03,MV-04,SBSVF-04,SVBSF-04,SBSVF-05,TIM-05,%%
  WBJSS-05,CIS-06,ODHIM-07,MHSF-08,KMPST-08,KHI-11,NCSOS-11,%%
  BMK-13,SP-15,DMPS-15, NCSOS-15,PV-19,PV-19-2}.  On the other hand,
multicomponent lattice AH systems have been much less considered in
two dimensions.  Beside their theoretical interest---their study allows us 
to deepen our understanding of statistical field theories with gauge
symmetries---2D lattice AH models may turn out to be relevant for
actual physical systems as well, as they may emerge in particular
regimes of low-dimensional condensed-matter systems, for example in
systems of ultracold atoms in optical lattices~\cite{BMTUZ-15}.

A lattice formulation of the AH model on a square lattice is obtained 
by considering  $N$-dimensional complex unit vector ${\bm z}_{\bm x}$ 
defined on the sites of the lattice
(they satisfy  $\bar{\bm z}_{\bm x} \cdot {\bm z}_{\bm x}=1$),
U(1) link variables 
$\lambda_{{\bm x},\mu}\equiv e^{i\theta_{\bm x}}$, and
the Hamiltonian
\begin{eqnarray}
H &=& - J N \sum_{{\bm x}, \mu}
\left( \bar{\bm{z}}_{\bm x} 
\cdot \lambda_{{\bm x},\mu}\, {\bm z}_{{\bm x}+\hat\mu} 
+ {\rm c.c.}\right)
\label{gllf}\\
&-& \gamma \sum_{{\bm x},\mu>\nu} 
\left(
\lambda_{{\bm x},{\mu}} \,\lambda_{{\bm x}+\hat{\mu},{\nu}} 
\,\bar{\lambda}_{{\bm x}+\hat{\nu},{\mu}}  
  \,\bar{\lambda}_{{\bm x},{\nu}} 
+ {\rm c.c.}\right)\,,
\nonumber
\end{eqnarray}
where $\hat\mu=\hat{1},\hat{2}$ are unit vectors along the lattice
directions, the first sum runs over all lattice links, while the
second one runs over all plaquettes.  Note that we use here the
standard Wilson compact formulation of electrodynamics.  The partition
function of the system reads
\begin{equation}
Z = \sum_{\{{\bm z}\},\{\lambda\}} e^{-\beta H}\,,\qquad \beta\equiv
1/T\,.
\label{partfun}
\end{equation}
We set $J=1$ without loss of generality (the explicit factor of $N$ in
the first term of the Hamiltonian is a standard choice, useful when
one considers the large-$N$ limit).  One can easily verify that the AH
model (\ref{gllf}) is invariant under a local U(1) gauge symmetry
${\bm z}_{\bm x} \to e^{i\Lambda_{\bm x}} {\bm z}_{\bm x}$,
$\lambda_{{\bm x},\mu} \to e^{i \Lambda_{\bm x}} \lambda_{{\bm x},\mu}
e^{-i \Lambda_{{\bm x}+\hat{\mu}}}$, and under a global SU($N$)
symmetry ${\bm z}_{\bm x} \to U {\bm z}_{\bm x}$ with $ U\in {\rm
  SU}(N)$.  The parameter $\gamma$ plays the role of inverse gauge
coupling.

For some particular values of $\gamma$ the AH model is equivalent to
simpler models.  In the limit $\gamma\to\infty$, the gauge link
variables are all equal to one modulo gauge transformations, and the
AH model becomes equivalent to the standard O($n$) vector model with
$n=2N$.  For $\gamma=0$, corresponding to the infinite gauge-coupling
limit, the AH model (\ref{gllf}) is a particular lattice formulation
of the CP$^{N-1}$
model~\cite{ZJ-book,RS-81,DHMNP-81,BL-81,CRV-92,CR-93}.  2D CP$^{N-1}$
models have been much studied in the literature, both analytically and
numerically, because they provide a theoretical laboratory to
understand some of the mechanisms of quantum field theories of
fundamental interactions. In particular, they share some notable
features with quantum chromodynamics (QCD), the theory of the hadronic
strong interactions, such as the asymptotic freedom and the so-called
$\theta$ dependence related to topology, see, e.g.,
Refs.~\cite{DVL-78,Witten-79,CR-92,VP-09}.  For both $\gamma=0$ and
$\gamma=\infty$, the lattice model shows a universal critical behavior
for $\beta\to\infty$. In this limit, the correlation length $\xi$
increases exponentially, $\xi\sim e^{c\beta}$.  Of course, the
critical behaviors for $\gamma=0$ and $\gamma\to\infty$ differ,
belonging to the universality class of the 2D CP$^{N-1}$ model for
$\gamma=0$ and to that of the 2D O($2N$) vector model for
$\gamma\to\infty$.

In this paper we investigate the phase diagram of 2D multicomponent
lattice AH models for generic values of the gauge parameter $\gamma$,
studying their critical behavior in the zero-temperature
limit. In three dimensions the AH model shows two
phases, separated by a transition where the global SU($N$) symmetry 
is broken and the gauge-invariant local
composite operator~\cite{PV-19-2}
\begin{equation}
Q_{{\bm x}}^{ab} = \bar{z}_{\bm x}^a z_{\bm x}^b - {1\over
  N}\delta^{ab}\,,
  \label{qdef}
\end{equation}
condenses.  In two dimensions, according to the Mermin-Wagner
theorem~\cite{MW-66}, such a condensation cannot occur.  Thus, we
expect the existence of a unique disordered phase. A critical behavior
only occurs for $\beta\to\infty$, as it happens in the 2D CP$^{N-1}$
model.

To investigate the critical behavior of 2D lattice AH models in the
zero-temperature limit, we present finite-size scaling (FSS) analyses
of numerical results obtained by Monte Carlo (MC) simulations.  The
asymptotic low-temperature behavior for generic finite values of
$\gamma$ turns out to be independent of $\gamma$, at least for
$\gamma$ not too negative. Therefore, it belongs to the 2D
universality class of the 2D CP$^{N-1}$ model. For any $\gamma$ we
observe an exponentially increase of the correlation length, the same
universal FSS curves, and the same dimensionless renormalization-group
(RG) invariant combinations of observables in the thermodynamic limit.
In the language of RG theory, the gauge coupling flows toward the
infinite gauge-coupling limit ($\gamma\to 0$), which represents the
stable fixed point of the RG flow.  In the zero gauge-coupling limit
($\gamma\to\infty$), the asymptotic large-$\beta$ behavior is expected
to change into that of the 2D O($2N$) $\sigma$ model, which
corresponds to an unstable fixed point of the RG flow.

The paper is organized as follows.  In Sec.~\ref{fsssec} we describe
the FSS framework which we employ to investigate the low-temperature
critical behavior.  In Sec.~\ref{infzelimits} we discuss two
particular limits of the lattice AH model: the infinite and zero
gauge-coupling limits.  In Sec.~\ref{results} we discuss the numerical
results for $N=2$ and $N=10$.  Finally, in Sec.~\ref{conclusions} we
draw our conclusions.

\section{Finite-size scaling}
\label{fsssec}

In this paper we investigate the nature of
the asymptotic large-$\beta$ behavior of the lattice
multicomponent AH model for different values of $\gamma$.
For this purpose we consider AH models on a  square lattice of linear
size $L$ with periodic boundary conditions.

We mostly focus on correlations of the gauge-invariant local
matrix variable $Q_{\bm x}^{ab}$ defined in
Eq.~(\ref{qdef}), which is hermitean and traceless.  Its two-point
correlation function is defined as
\begin{equation}
G({\bm x}-{\bm y}) = \langle {\rm Tr}\, Q_{\bm x} Q_{\bm y} \rangle\,,  
\label{gxyp}
\end{equation}
where the translation invariance of the system has been taken into
account.  The susceptibility and the correlation length are defined as
$\chi=\sum_{\bm x} G({\bm x})$ and
\begin{eqnarray}
\xi^2 \equiv  {1\over 4 \sin^2 (\pi/L)}
{\widetilde{G}({\bm 0}) - \widetilde{G}({\bm p}_m)\over 
\widetilde{G}({\bm p}_m)}\,,
\label{xidefpb}
\end{eqnarray}
where $\widetilde{G}({\bm p})=\sum_{{\bm x}} e^{i{\bm p}\cdot {\bm x}}
G({\bm x})$ is the Fourier transform of $G({\bm x})$, and ${\bm p}_m =
(2\pi/L,0)$.  We also consider
the Binder parameter defined as
\begin{equation}
U = {\langle \mu_2^2\rangle \over \langle \mu_2 \rangle^2} \,, \qquad
\mu_2 = 
\sum_{{\bm x},{\bm y}} {\rm Tr}\,Q_{\bm x} Q_{\bm y}\,.
\label{binderdef}
\end{equation}
To determine the universal features of the asymptotic behavior of the AH model,
we use a FSS approach
~\cite{FB-72,Barber-83,Privman-90,PHA-91,PV-02,GKMD-98}.
At finite-temperature continuous transitions the FSS limit is obtained by
taking $\beta\to \beta_c$ and $L\to\infty$ keeping $X \equiv
(\beta-\beta_c)L^{1/\nu}$ fixed, where $\beta_c$ is the inverse
critical temperature and $\nu$ is the correlation-length exponent.
Any RG invariant quantity $R$, such as the ratio
\begin{equation}
R_\xi\equiv \xi/L
\label{rxidef}
\end{equation}
and the Binder parameter $U$, is expected to asymptotically behave as
$R(\beta,L) = f_R(X)+O(L^{-\omega})$, where $\omega$ is a universal exponent. 
The scaling function $f_R(X)$ is
universal apart from a trivial normalization of its argument; it only
depends on the shape of the lattice and on the boundary conditions.
Since $R_\xi$ is generally
monotonic, we can also write~\cite{Barber-83,Privman-90,PHA-91,PV-02}, 
\begin{equation}
R(\beta,L) = F_R(R_\xi) + O(L^{-\omega}),
\label{r12sca}
\end{equation}
where $F_R$ is a universal scaling function.
Eq.~(\ref{r12sca}) is particularly convenient, as it allows a direct
check of universality, without the need of tuning any parameter. Moreover, it 
applies directly, without any change, to two-dimensional asymptotically
free models, like the 2D CP$^{N-1}$ model and the 2D O($N$) nonlinear $\sigma$
model~\cite{ZJ-book}, in which 
a critical behavior is only obtained in the limit $\beta \to \infty$,
see Refs.~\cite{LWW-91,Kim-93,CEFPS-95,CP-98,BNW-10} and references therein. 
In this
case, scaling corrections decay
as $L^{-2} \log^p L$, where $p$ cannot be determined in perturbation theory 
(see Ref.~\cite{CP-98} for a discussion in the O($N$) model).

In the following, we will consider the finite-size behavior of the 
Binder parameter $U$, which varies between \cite{PV-19}
\begin{eqnarray}
{\rm lim}_{R_\xi\to 0} \;U = {N^2+1\over N^2-1}\,,\qquad
{\rm lim}_{R_\xi\to \infty}\; U = 1\,. \label{uextr}
\end{eqnarray}
To understand the asymptotic critical 
behavior of different models we will compare the behavior of $U$ as 
a function of $R_\xi$: If two models belong to the same
universality class, the Binder parameter $U$ must satisfy the 
FSS relation
(\ref{r12sca}) with the same asymptotic curve $F_U(R_\xi)$.
This is expected to emerge when
comparing results for the lattice CP$^1$ model, obtained by setting
$\gamma=0$ in the $N=2$ AH model (\ref{gllf}), and the standard 
O(3) vector model.
This will be explicitly shown in
Sec.~\ref{n2comp}.
It will also allow us to
show that the $N$-component lattice AH model has the same FSS behavior as that
of the CP$^{N-1}$ model, independently of $\gamma$, demonstrating that
their asymptotic large-$\beta$ critical behavior belongs to the
same 2D universality class, characterized by a local U(1) gauge
invariance and a global SU($N$) symmetry. Universality also implies 
that all dimensionless RG invariant quantities
have the same asymptotic
large-$\beta$ behavior, both in the thermodynamic and in the FSS limit.

\section{Infinite and zero gauge coupling}
\label{infzelimits}

In this section we discuss two particular limits of the lattice AH
model: the infinite and zero gauge-coupling limits, corresponding to
$\gamma=0$ and $\gamma\to\infty$ respectively.

For $\gamma = 0$, the AH model corresponds to a
lattice formulation of the CP$^{N-1}$ model.  Indeed, for
$\gamma=0$ one can integrate out the link variables $\lambda_{{\bm
    x},\mu}$ in the partition function, obtaining
\begin{equation}
Z = \sum_{\{z\}}
\prod_{{\bm x},\mu} 
I_0\left(2\beta N |\bar{\bm z}_{\bm x} \cdot {\bm z}_{{\bm
    x}+\hat\mu}|\right),
\label{partzint}
\end{equation}
where $I_0(x)$ is a modified Bessel function.  The corresponding
effective Hamiltonian reads
\begin{equation}
H_{\rm eff} = - \beta^{-1} 
\sum_{{\bm x},\mu}  
\ln I_0\left(2\beta N |\bar{\bm{z}}_{\bm x} \cdot {\bm
      z}_{{\bm x}+\hat\mu}|\right)\,.
\label{hlaeff}
\end{equation}
Taking into account that $I_0(x) = 1 + x^2/4 +
O(x^4)$, one recovers the CP$^{N-1}$ model in the continuum
limit, i.e., the quantum field theory defined as
\begin{eqnarray}
&&{\cal Z} = \int [d{\bm z}] \ \exp \left[{-\int d{\bm x} {\cal
        L}({\bm z})}\right] \,, \nonumber \\ 
&& {\cal L} = {1\over 2 g}
  \overline{D_{\mu}{\bm z}}\cdot D_\mu {\bm z}\,, \qquad D_\mu =
  \partial_\mu + i A_\mu\,, \qquad
\label{contham}
\end{eqnarray}
where $A_\mu = i\bar{{\bm z}}\cdot \partial_\mu {\bm z}$ is a
composite gauge field.  For $N=2$ the CP$^1$ field
theory is locally isomorphic to the O(3) non-linear $\sigma$ model. 
The mapping is obtained by identifying 
the three-component real vector $s_{\bm x}^a$ with the combination
$\sum_{ij} \bar{z}_{\bm x}^i \sigma_{ij}^{a} z_{\bm x}^j$, where
$a=1,2,3$ and $\sigma^{a}$ are the Pauli matrices.

In two dimensions, the 
lattice CP$^{N-1}$ model is asymptotically free, a property it shares 
with QCD, the theory of strong interactions.
It does not show a finite-temperature transition, but nonetheless it shows 
a universal behavior for $\beta\to \infty$. In this limit the correlation
length behaves as (see, e.g., Ref.~\cite{CRV-92})
\begin{equation}
\xi =  A (2\pi \beta)^{-2/N} e^{2\pi\beta} [1 + O(\beta^{-1})]\,,
\label{xibeta}
\end{equation}
where $A$ is a model-dependent constant.

In the limit $\gamma\to\infty$, the gauge link variables are all equal
to one, modulo gauge transformations. Therefore, the lattice AH model
(\ref{gllf}) can be exactly mapped onto a lattice O($n$) vector model
with $n=2N$, with standard Hamiltonian
\begin{equation}
H_{{\rm O}(n)} = - \sum_{{\bm x}, \mu} s_{\bm x} \cdot s_{{\bm
    x}+\hat{\mu}},
\label{onmodel}
\end{equation}
where $s_{\bm x}$ are $n$-component real vectors satisfying $s_{\bm
  x}\cdot s_{\bm x}=1$.  We recall that in two dimensions, also O$(n)$
vector models are asymptotically free: A critical behavior is only
obtained for $\beta\to\infty$. In this limit, the correlation length
increases exponentially \cite{ZJ-book}:
\begin{equation}
\xi =  B (2 \pi \hat{\beta})^{-1/(n-2)} e^{2 \pi \hat{\beta}} ,
\end{equation}
where $B$ is a model-dependent constant and $\hat{\beta} =
\beta/(n-2)$.  For the lattice O($n$) vector model, one can define a
second-moment correlation using Eq.~(\ref{xidefpb}) and the two-point
function
\begin{equation}
G_{{\rm O}(n)} ({\bm x},{\bm y})\equiv 
\langle s_{\bm x}\cdot s_{\bm y}\rangle\,.
\label{gon}
\end{equation}
The Binder parameter is defined analogously as
\begin{equation}
U_{{\rm O}(n)} = {\langle m_2^2\rangle \over \langle m_2
  \rangle^2}\,,\qquad m_2 = \sum_{{\bm x},{\bm y}} G_{{\rm O}(n)}({\bm
  x},{\bm y})\,.
\label{uon}
\end{equation}

Although the AH model is mapped onto the O($n$) model for $\gamma\to
\infty$, one should note that the mapping between AH and O($n$)
observables is not trivial. The correlation length and the Binder
parameter defined using Eqs.~(\ref{gon}) and (\ref{uon}) in the O($n$)
model do not correspond to those defined in the AH model.  The correct
correspondence is discussed in Ref.~\cite{PV-19-2}, where the same
issue was discussed in three dimensions.

\section{Numerical results in the zero-temperature limit}
\label{results}

To identify the universal features of the large-$\beta$ behavior of
the AH model, we have performed MC simulations for $N=2$ and 10 and
several values of $\gamma$. The linearity of Hamiltonian (\ref{gllf})
with respect to each lattice variable allows us to employ an
overrelaxed algorithm for the updating of the lattice configurations.
It consists in a stochastic mixing of microcanonical and standard
Metropolis updates.  To update each lattice variable, we randomly
choose either a standard Metropolis update, which ensures ergodicity,
or a microcanonical move, which is more efficient than the Metropolis
one but does not change the energy. On average, we perform three/four
microcanonical updates for every Metropolis proposal.  In the
Metropolis update, changes are tuned so that the acceptance is
1/3. The same algorithm was used in Ref.~\cite{PV-19-2} for 3D
systems.

\subsection{The lattice AH model with  $N=2$}
\label{n2comp}

We first consider the $N=2$ AH model. We will show that the
critical behavior in the zero-temperature limit belongs to the
universality class of the 2D O(3) nonlinear $\sigma$ model
independently of $\gamma$, at least for $\gamma$ not too negative.

\begin{figure}[tbp]
\includegraphics*[scale=\graphicscale]{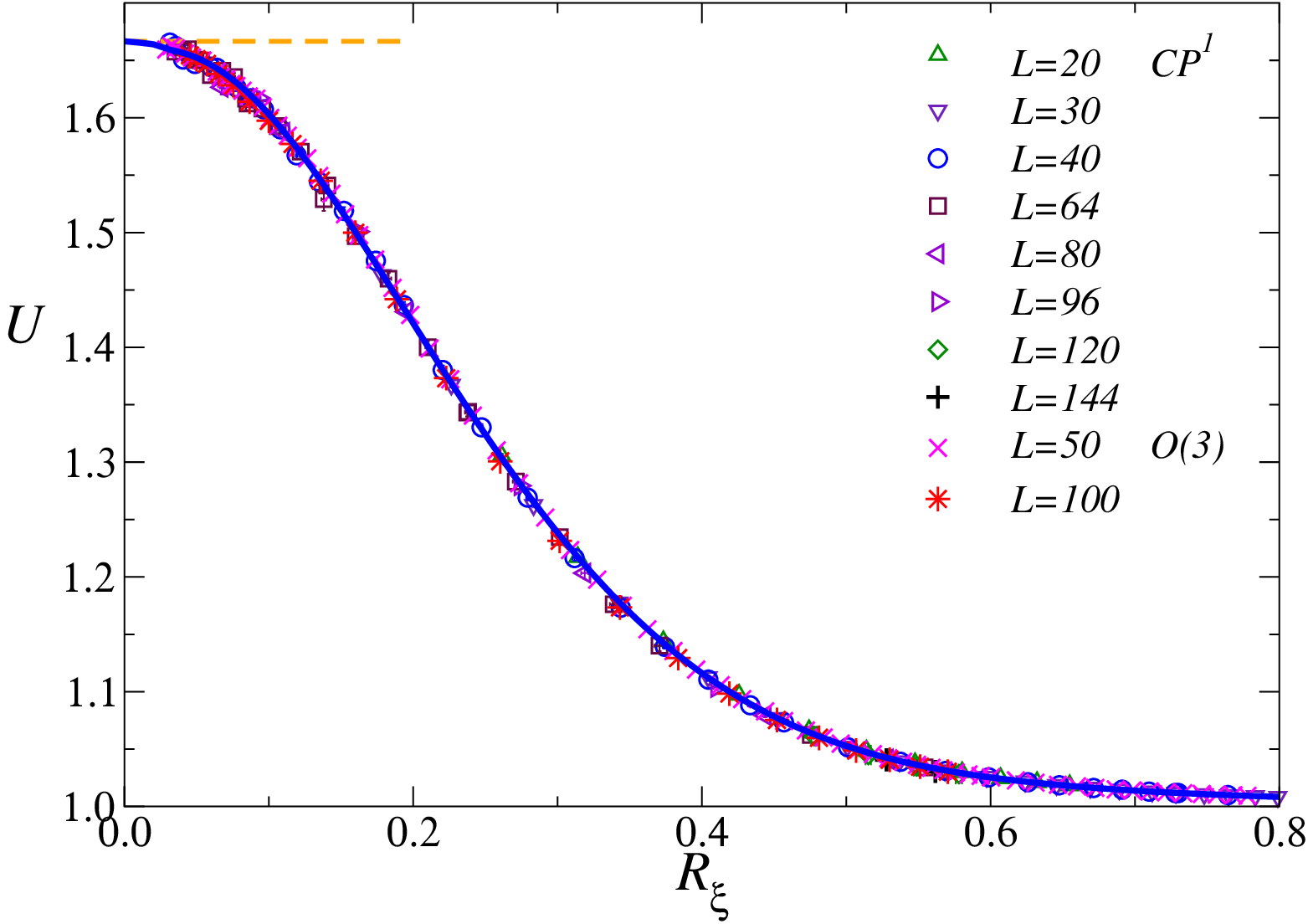}
\caption{Estimates of $U$ versus $R_\xi$ for the CP$^1$ model ($N=2$
  AH model with $\gamma=0$), up to $L=144$, and for the standard
  lattice O(3) $\sigma$ model (up to $L=100$). The data appear to
  converge to the same FSS curve.  The horizontal dashed line
  corresponds to the asymptotic value $U=5/3$ for $R_\xi\to 0$.  The
  full line represents an interpolation of the O(3) data, which
  provides the universal FSS curve with an accuracy of few per mille
  for $R_\xi\lesssim 0.8$~\cite{footnote-interpolation}.}
\label{bi-rxi-cp1}
\end{figure}

To begin with, we present results for $\gamma=0$, i.e., for the CP$^1$
model.  As discussed in Sec.~\ref{infzelimits}, the CP$^1$ lattice
model can be mapped onto a lattice O(3) $\sigma$ model, although
different from the standard one reported in Eq.~(\ref{onmodel}), see
e.g. Ref.~\cite{CRV-92}. It is easy to verify that, under this
mapping, the two-point function, the second-moment correlation length
$\xi$, and the Binder parameter $U$ of the CP$^1$ model, defined in
Eqs.~(\ref{gxyp}), (\ref{xidefpb}) and (\ref{binderdef}),
respectively, correspond to those of the O(3) vector model
(\ref{onmodel}) defined in Eqs.~(\ref{gon}) and (\ref{uon}). We thus
expect the large-$\beta$ critical behavior of the AH model for
$\gamma=0$ to have the same universal features as that of the O(3)
$\sigma$ model.  In Fig.~\ref{bi-rxi-cp1} we report the Binder
parameter $U$ versus $R_\xi$, up to $L=144$. These results are
compared with those obtained by cluster \cite{Wolff-89} simulations of
the standard O(3) $\sigma$ model (\ref{onmodel}) up to $L=100$.  As
expected, the data of $U$ versus $R_\xi$ for the two lattice models
converge towards the same scaling curve with increasing $L$,
conferming that they belong to the same 2D universality class.

\begin{figure}[tbp]
\includegraphics*[scale=\graphicscale]{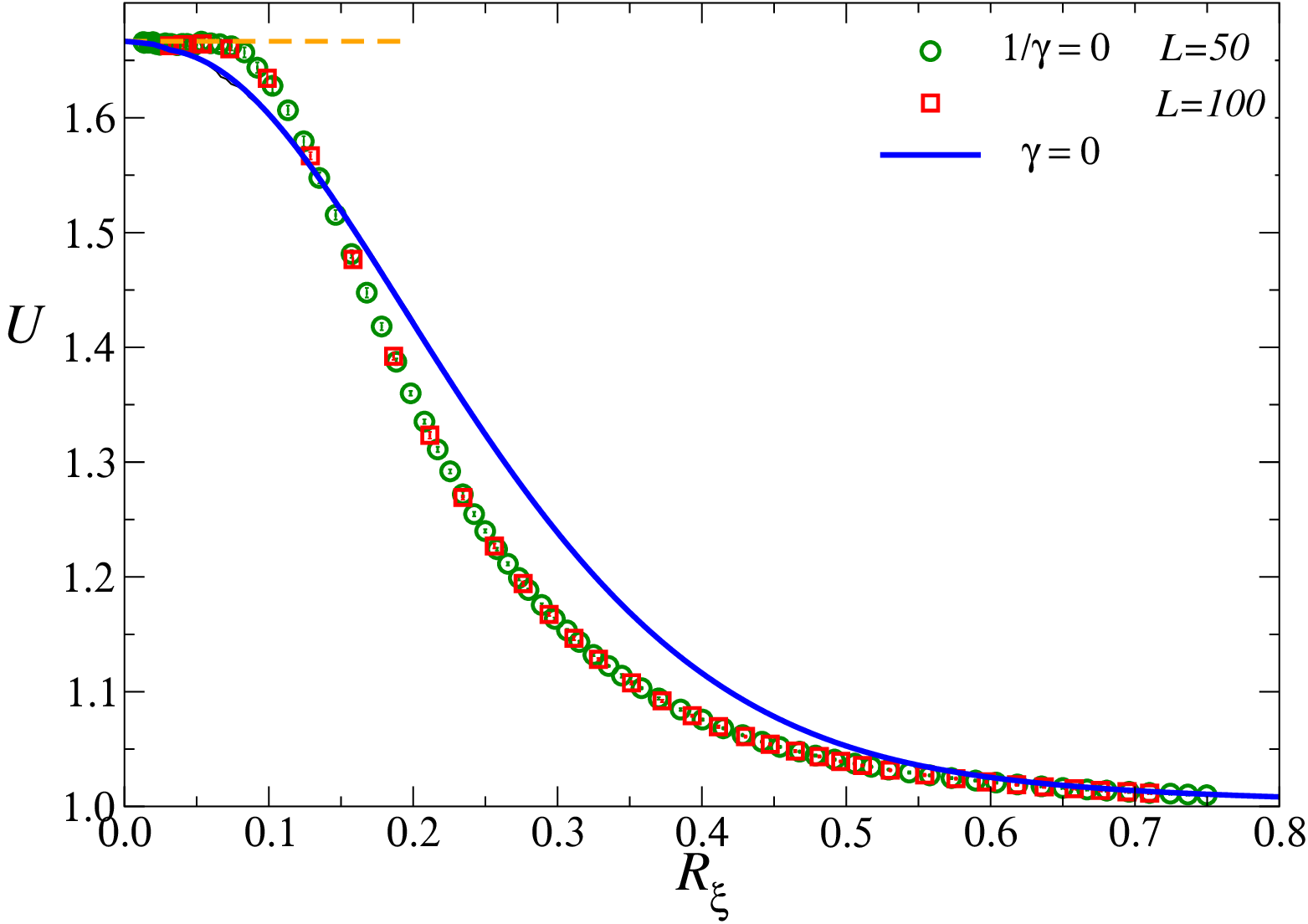}
\caption{Plot of $U$ versus $R_\xi$ for $\gamma\to\infty$, as computed
  in the O(4) vector model.  The horizontal dashed line correspomds to
  the asymptotic value $U=5/3$ for $R_\xi\to 0$.  For comparison we
  also report the O(3)/CP$^1$ FSS curve ($\gamma=0$), obtained from
  the interpolation~\cite{footnote-interpolation} of the O(3) data
  shown in Fig.~\ref{bi-rxi-cp1}. }
\label{bi-rxi-o4}
\end{figure}

In the limit $\gamma\to\infty$, the $N=2$ AH model becomes equivalent
to the standard lattice O(4) vector model defined in
Eq.~(\ref{onmodel}).  Therefore, for $\gamma\to\infty$ the asymptotic
large-$\beta$ behavior should correspond to that of the O(4) vector
model.  In Fig.~\ref{bi-rxi-o4} we show the AH Binder parameter
defined in Eq.~(\ref{binderdef}) as computed in the O(4) model (as
already mentioned, this is not the usual O(4) Binder parameter, see
Ref.~\cite{PV-19-2} for details).  The curves for $\gamma=0$ and
$\gamma=\infty$ are clearly different, although they both converge to
$U=5/3$ and $U=1$ for $R_\xi\to 0$ and $R_\xi\to\infty$, respectively.

\begin{figure}[tbp]
\includegraphics*[scale=\graphicscale]{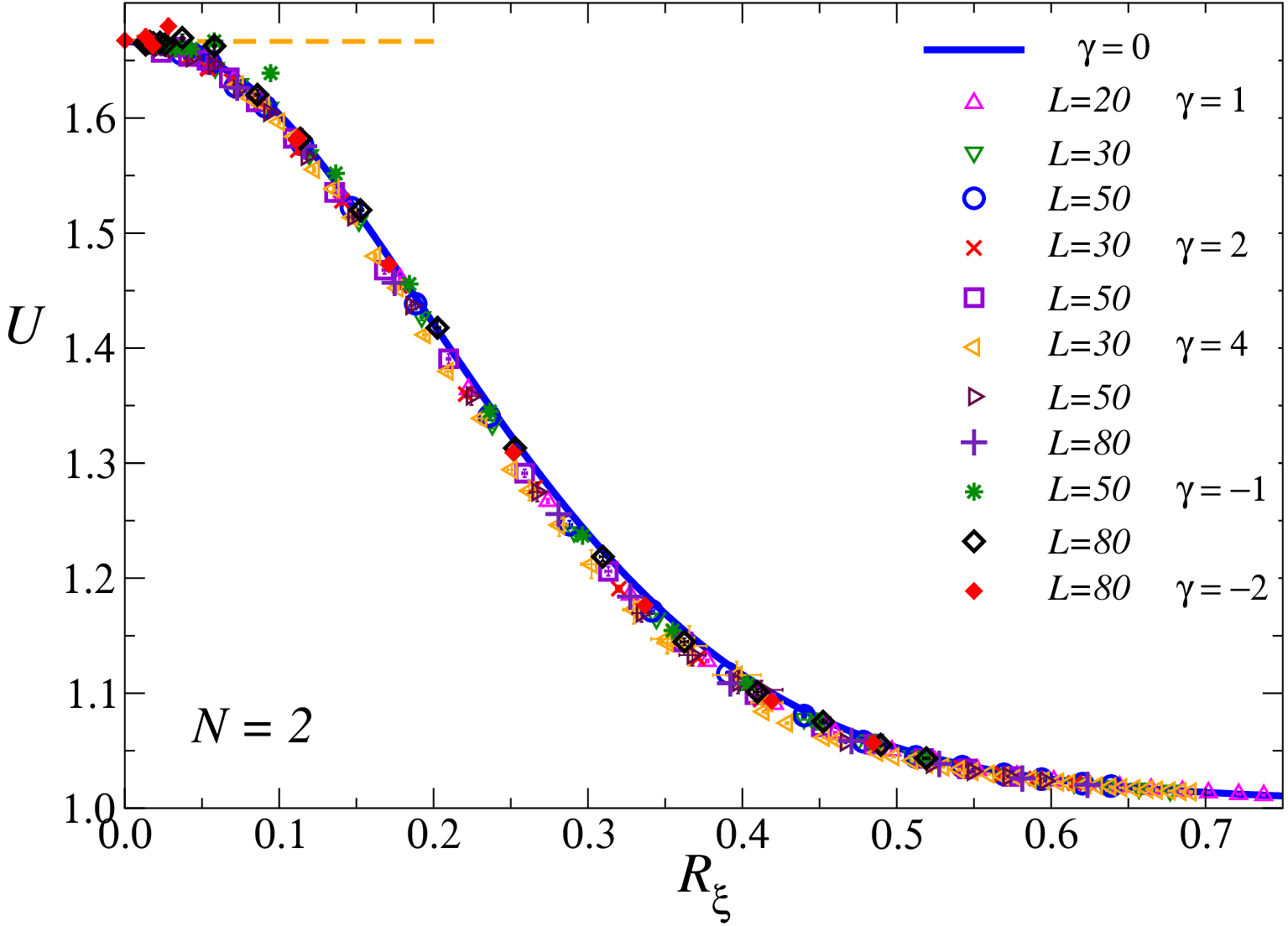}
\caption{Plot of $U$ versus $R_\xi$ for the $N=2$ AH model for several
  lattice sizes and values of $\gamma$ in the range $[-2,4]$.  The
  data appear to approach the universal curve associated with the O(3)
  universality class (full line, $\gamma=0$), obtained by
  interpolating~\cite{footnote-interpolation} the O(3) results shown
  in Fig.~\ref{bi-rxi-cp1}.  }
\label{bi-rxi-N2}
\end{figure}

We now consider the lattice AH model for finite nonzero values of
$\gamma$.  Results for $\gamma\in [-2,4]$ are shown in
Fig.~\ref{bi-rxi-N2}.  In all cases the data of $U$ versus $R_\xi$
appear to approach the curve of the O(3) vector model.  These results
show that in a wide interval of values of $\gamma$ around zero, the
$N=2$ lattice AH model has the same critical behavior as the O(3)
vector model.  Differences decay rapidly with $L$, consistently with
the expected $O(L^{-2})$ approach.

\begin{figure}[tbp]
\includegraphics*[scale=\graphicscale]{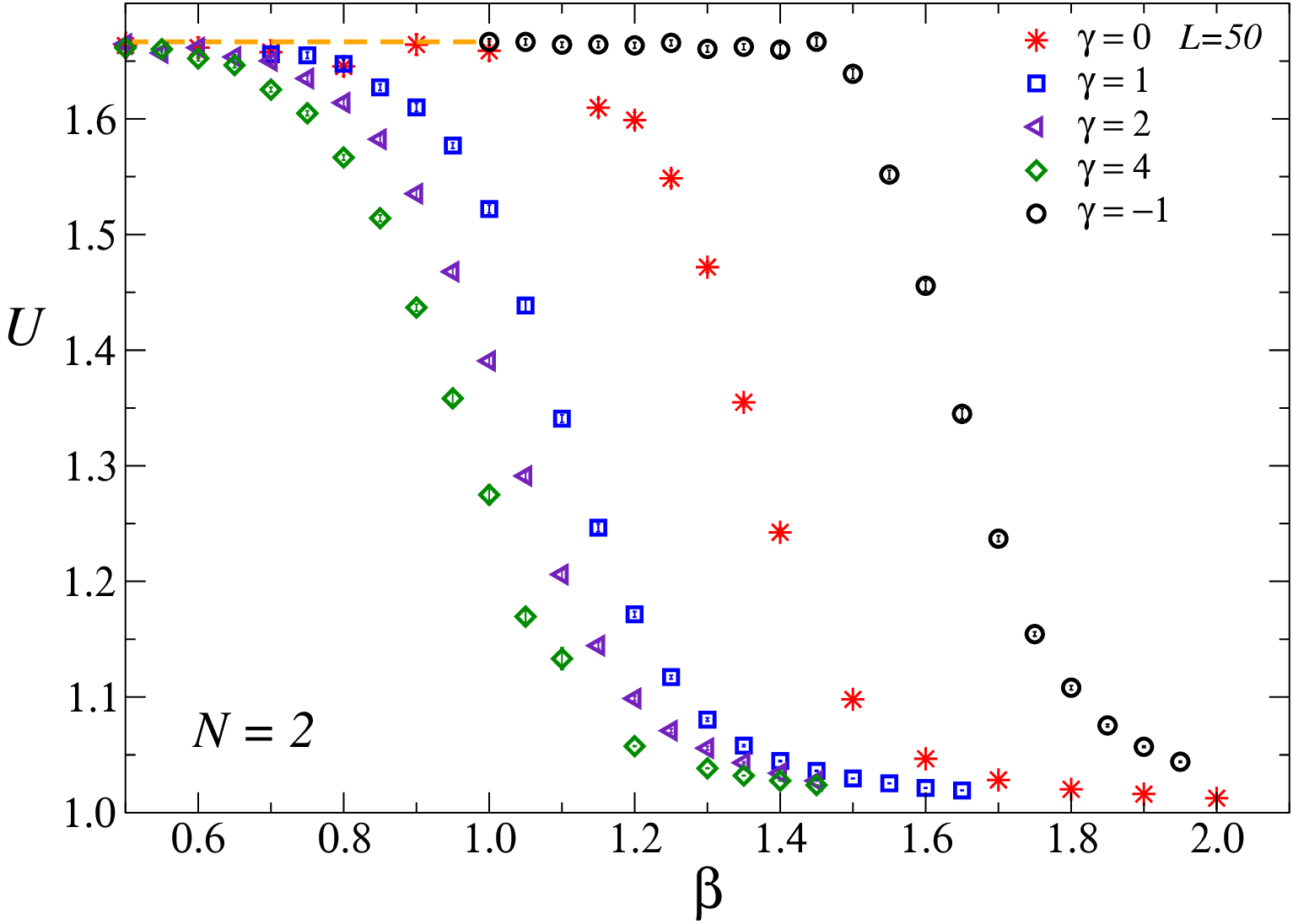}
\caption{Estimates of $U$ versus $\beta$ for the $N=2$ AH model for
  several values of $\gamma$, in the range $-1\le \gamma\le
  4$. Results for $L=50$. }
\label{bi-beta-N2}
\end{figure}

Notice that the universality of the curves of $U$ versus $R_\xi$ for
different values of $\gamma$ is not trivial, since the dependence on
$\beta$ significantly changes when changing $\gamma$, as demonstrated
in Fig.~\ref{bi-beta-N2}. In the language of asymptotically free
theories, see, e.g., Ref.~\cite{CRV-92}, the 2D lattice AH models with
different values of $\gamma$ are variant actions of the 2D CP$^{N-1}$
model with $\Lambda$ parameters that significantly depend on $\gamma$.

On the basis of the previous results, we conjecture that the $N=2$ AH
model has the same critical behavior as the O(3) vector model for any
positive finite $\gamma$.  The same conjecture may not hold for all
negative values.  Indeed, for $\gamma$ large and negative one would
obtain a fully frustrated model, which might have a critical behavior
distinctly different from that of a ferromagnetic model. This indeed
occurs in the fully frustrated XY model \cite{FFXY}, which corresponds
to the $N=1$ AH model in the $\gamma\to-\infty$ limit.  In any case,
our results show that, at least for $\gamma> -2$, the behavior is the
same as that for $\gamma\ge 0$.  In the RG language, these results
imply that, at least for $\gamma$ not too negative, the RG flow of the
2D $N=2$ lattice AH models is controlled by the stable infinite
gauge-coupling fixed point, belonging to the 2D O(3) vector
universality class.

\begin{figure}[tbp]
\includegraphics*[scale=\graphicscale]{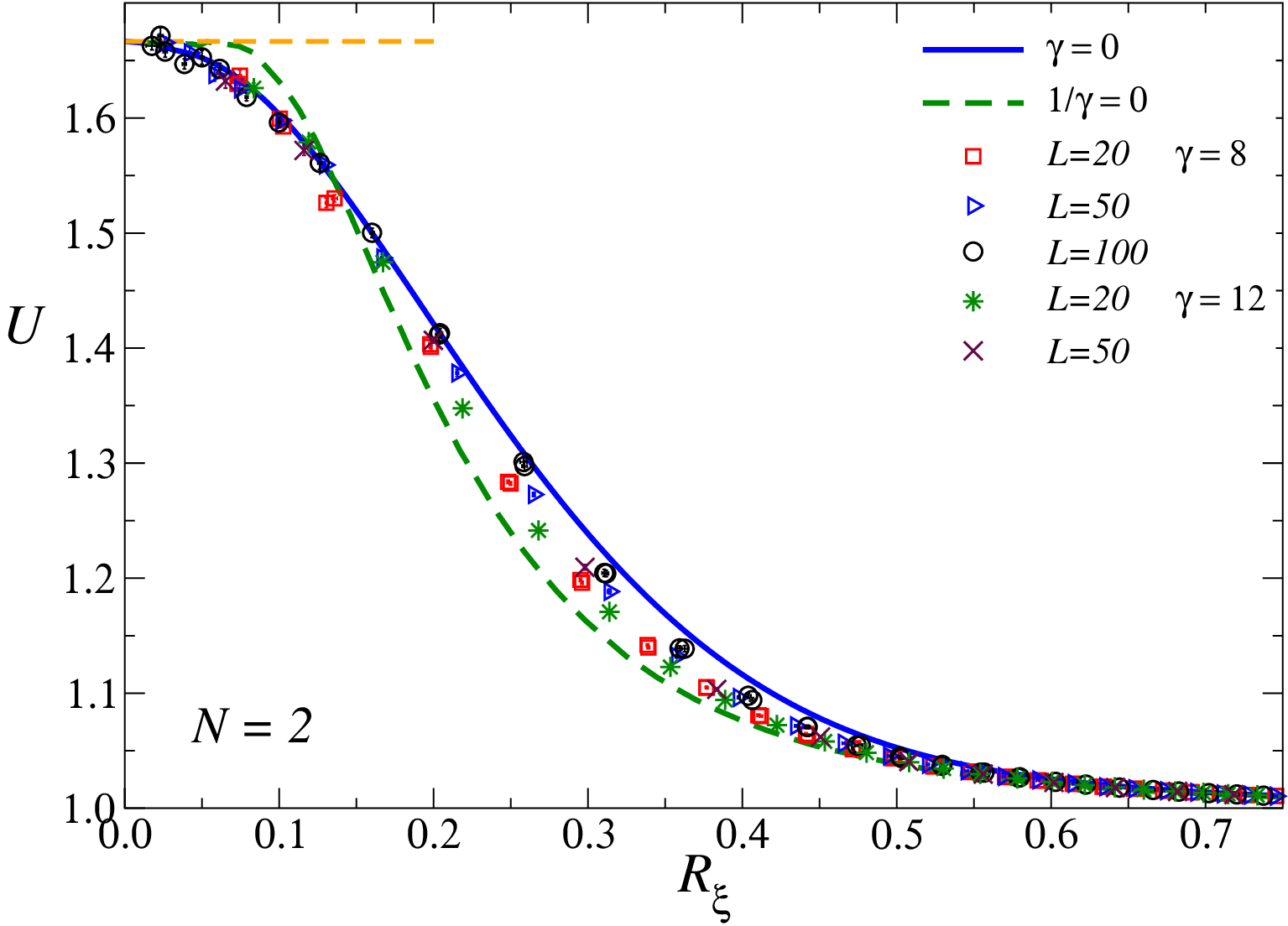}
\caption{Estimates of $U$ versus $R_\xi$ for the $N=2$ AH model with
  $\gamma=8,\,12$.  Data are compared with the FSS curves of the
  lattice AH model in the limit $\gamma\to\infty$ (dashed line) and
  $\gamma=0$ (full line). The data for the smallest lattice sizes are
  close to the $\gamma\to\infty$ FSS curve, then they approach the
  $\gamma=0$ curve with increasing $L$. }
\label{bi-rxi-N2-laga}
\end{figure}

Since the AH model becomes equivalent to the O(4) vector model for
$\gamma \to\infty$, we expect significant crossover phenomena for
large values of $\gamma$.  This is confirmed by the MC data for
$\gamma=8$ and 12 reported in Fig.~\ref{bi-rxi-N2-laga}.  They provide
evidence of the predicted crossover behavior. For small values of $L$
the data are close to the O(4) ($\gamma=\infty$) scaling curve, and
then move systematically towards the asymptotic O(3) curve.

\begin{figure}[tbp]
\includegraphics*[scale=\graphicscale]{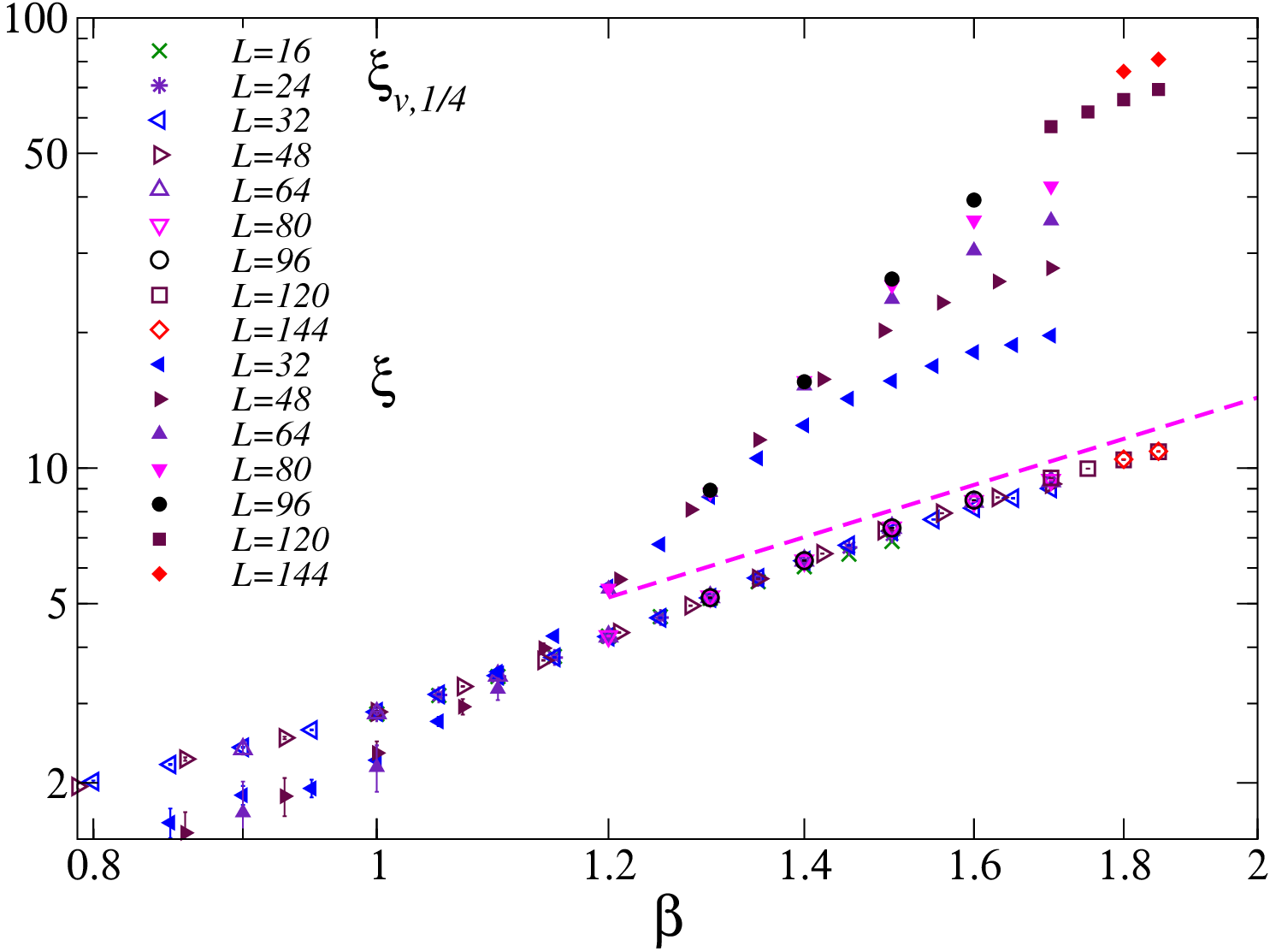}
\caption{ Estimates of the vector correlation length $\xi_{v,1/4}$,
  defined in Eq.~(\ref{xigdef}), and of the correlation length $\xi$,
  for the lattice CP$^1$ model ($\gamma=0$). The dashed line
  corresponds to a $\beta^2$ behavior.  }
\label{xigl-beta}
\end{figure}

It is also worth discussing the behavior of gauge-invariant
correlations of the vector variable ${\bm z}_x$, such as $\langle
\bar{\bm z}_{\bm x}\cdot {\bm z}_{\bm y} \prod_{l \in \cal C}
\lambda_l \rangle$, where the product extends over the link variables
that belong to a lattice path $\cal C$ connecting points $\bm x$ and
$\bm y$.  In particular, we consider correlations between points that
belong to lattice straight lines.  We select a generic lattice
direction $\hat{\mu}$ and define the vector correlation function
\begin{eqnarray}
G_v(\ell,\beta)={1\over L^2} \sum_{{\bm x}} \hbox{Re}\left\langle
\bar{\bm z}_{\bm x}\cdot {\bm z}_{\bm y}
\prod_{n=0}^{\ell-1} \lambda_{{\bm x}+n \hat{\mu},\mu}
\right\rangle\,,
   \label{Gd}
\end{eqnarray}
where ${\bm y}={{\bm x}+\ell \hat{\mu}}$ and all coordinates should be
taken modulo $L$, because of the periodic boundary conditions. We
average over all lattice sites ${\bm x}$ exploiting the translation
invariance of systems with periodic boundary conditions.  Note that
the correlation length $G_v$ is not periodic, i.e., it does not
satisfy $G_v(\ell+L,\beta) = G_v(\ell,\beta)$, because of the string
of gauge fields between the spatial points.

For sufficiently large lattice sizes and relatively small distances
$\ell$, $G_v$ has an approximately exponentially decaying behavior.
By assuming that $G_v\sim e^{-\ell/\xi_g}$, one can define a vector
length scale as
\begin{eqnarray}
&& \xi_{v,X} \equiv  {L X/2\over  \ln[
 G_v(L X/2,\beta)/G_v(L X,\beta)]}\,,
 \label{xigdef}
\end{eqnarray} 
where $0 \le X < 1$ is a parameter.  We have measured $\xi_{v,X}$ for
$X=1/4$ (other values of $X$ give similar results) and $\gamma=0$. At
variance with what happens with $\xi$, $\xi_{v,X}$ shows very small
size corrections. The asymptotic infinite-volume results are obtained
on quite small lattices, even when $\xi$ is still of the order of
$L$. Apparently, gauge modes and gauge-invariant modes encoded in
$\xi$ are little correlated.  The results, reported in
Fig.~\ref{xigl-beta}, are definitely consistent with a simple
power-law behavior: apparently they are compatible with $\xi_v\sim
\beta^2$. Therefore, for $\beta \to \infty$, $\xi_v$ increases much
slower than $\xi$, which instead increases exponentially, see
Fig.~\ref{xigl-beta}.  These results imply
\begin{equation}
\xi_v/\xi\to 0 
\label{xigxi}
\end{equation}
exponentially, for $\beta\to \infty$.  Therefore, vector and gauge
modes are irrelevant in the continuum limit, explaining why $\gamma$
is an irrelevant parameter in the theory, i.e., why it does not change
the universality class of the model.

\subsection{The lattice AH model with  $N=10$}
\label{n10comp}

\begin{figure}[tbp]
\includegraphics*[scale=\graphicscale]{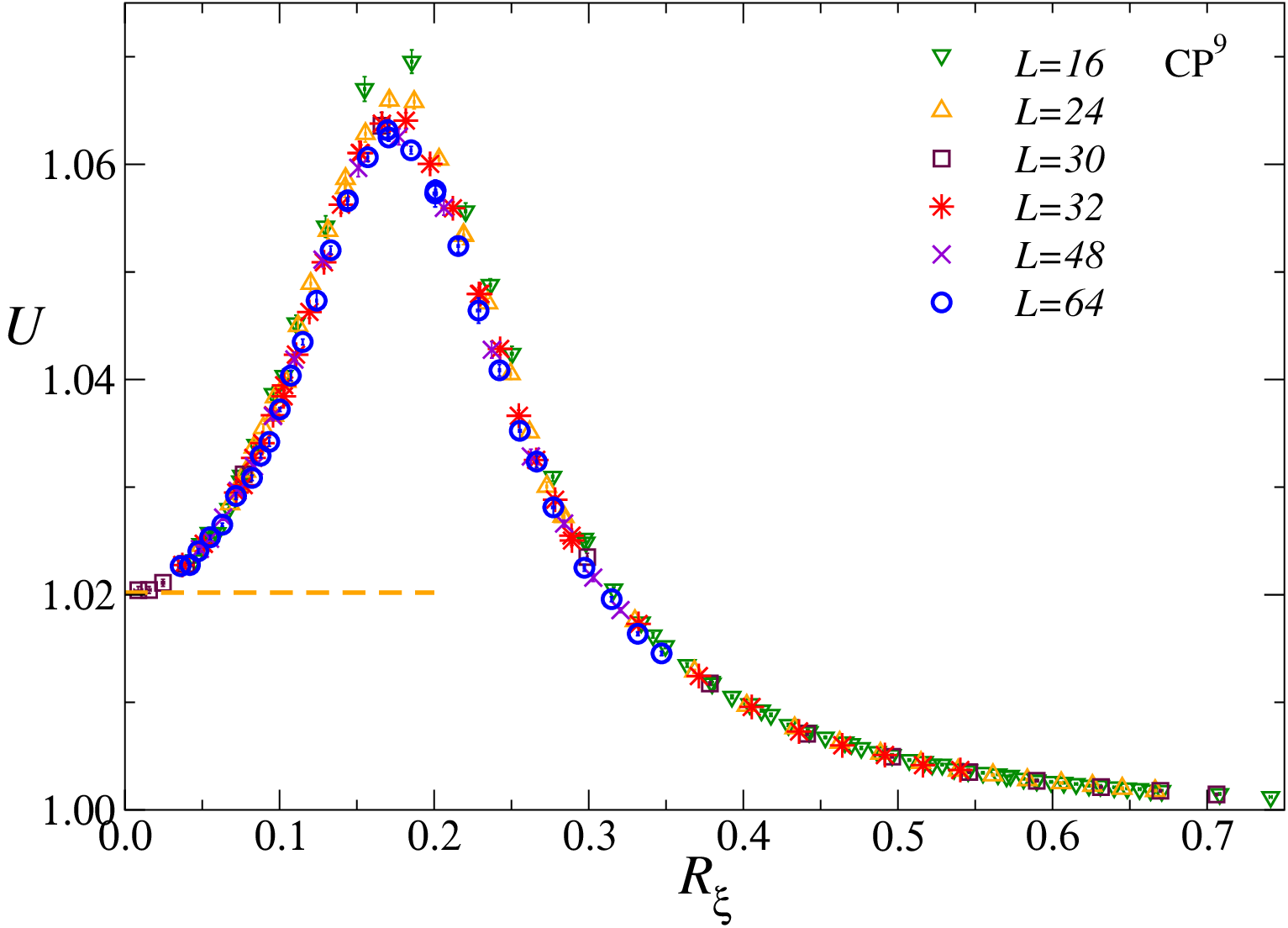}
\caption{Estimates of $U$ versus $R_\xi$ for the CP$^9$ model.  The
  horizontal dashed line corresponds to the asymptotic value
  $U=101/99$ for $R_\xi\to 0$. The data appear to converge to a
  universal FSS curve.}
\label{bi-rxi-cp9}
\end{figure}

\begin{figure}[tbp]
\includegraphics*[scale=\graphicscale]{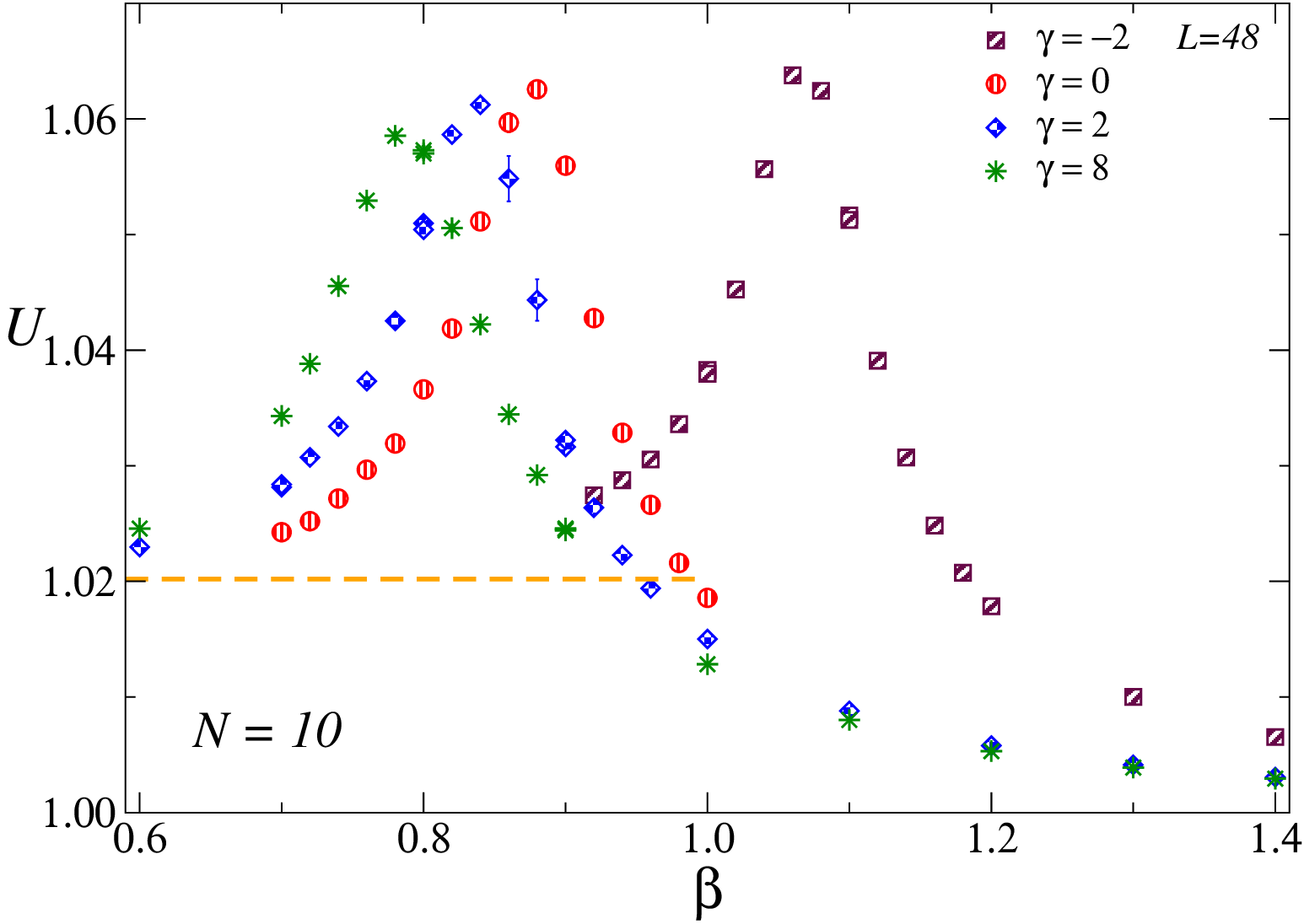}
\includegraphics*[scale=\graphicscale]{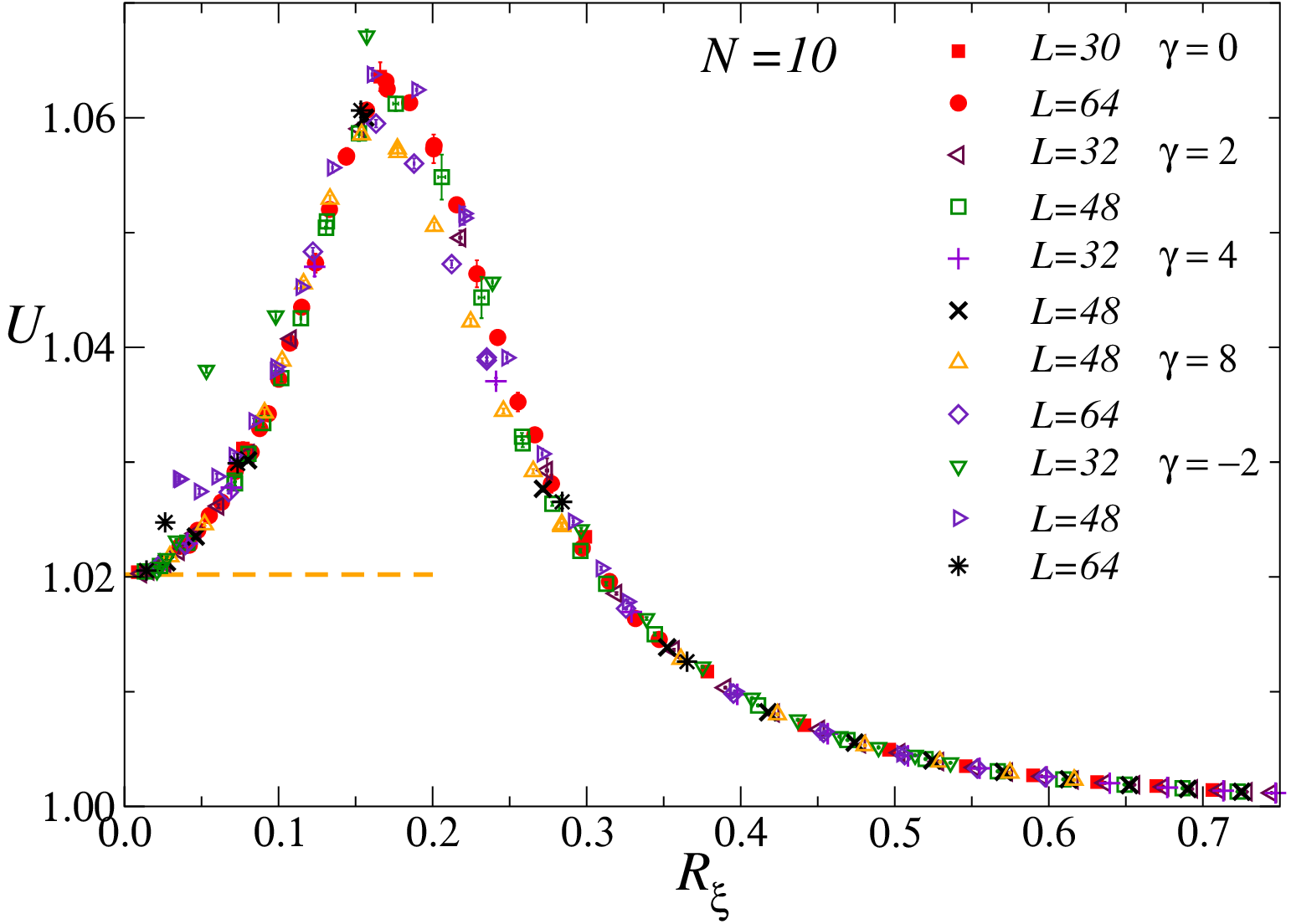}
\caption{Estimates of $U$ versus $\beta$ (top) and versus $R_\xi$
  (bottom) for the $N=10$ AH model for several lattice sizes and 
  values of $\gamma$ in the range $\gamma\in [-2,8]$.  With
  increasing $L$, the curves of $U$ vs $R_\xi$ appear to converge
  toward the universal CP$^9$ curve (AH model with $\gamma=0$). }
\label{bi-rxi-n10}
\end{figure}

We now consider the lattice AH model with $N=10$.
Fig.~\ref{bi-rxi-cp9} shows $U$ versus $R_\xi$ for the model with
$\gamma=0$, which is a particular lattice formulation of the CP$^9$
model.  The data converge to a universal FSS curve. At variance with
the CP$^1$ model, the FSS curve is nonmonotonic. It shows a pronounced
peak, which is related to the presence of bound states in the
large-$N$ limit, whose physical size increases as $N^{1/3}$, as
already noted in Ref.~\cite{RV-93}.

As in the $N=2$ case, the numerical results show that the asymptotic
FSS curve is independent of $\gamma$.  This is clearly shown by the
data reported in Fig.~\ref{bi-rxi-n10}: The data fall on top of the
CP$^9$ results for all values of $\gamma$ in $[-2,8]$. Note that this
agreement is not trivial, as the $\beta$ dependence of $U$ varies
significantly when changing $\gamma$.  We note again that larger
scaling corrections are observed with increasing $\gamma$. But this is
again expected, due to the fact that in the large-$\gamma$ limit the
model becomes equivalent to a lattice O(20) vector model, see
Sec.~\ref{infzelimits}.

These results show that the RG flow of the 2D $N=10$ lattice AH model
is generally controlled by the stable infinite gauge-coupling fixed
point, corresponding to the 2D CP$^9$ theory, as it occurs for $N=2$.
For large values of $\gamma$, crossover phenomena are again
expected, due to the unstable zero gauge-coupling fixed point
corresponding to a 2D O(20) vector model.

\section{Conclusions}
\label{conclusions}

We have investigated the critical behavior of the 2D multicomponent
lattice AH model with Hamiltonian ~(\ref{gllf}). It represents a 2D
lattice U(1) gauge theory coupled to an $N$-component complex scalar
field.  Beside the U(1) gauge invariance, it is invariant under a
global SU($N$) symmetry. In agreement with the Mermin-Wagner theorem,
it does not show an ordered phase where the global SU($N$) symmetry is
spontaneously broken.  Therefore, no finite-temperature transition
occurs and, for any fixed $\gamma$, a critical behavior is only
observed for $\beta\to\infty$.

\begin{figure}[tbp]
\includegraphics*[scale=\graphicscale]{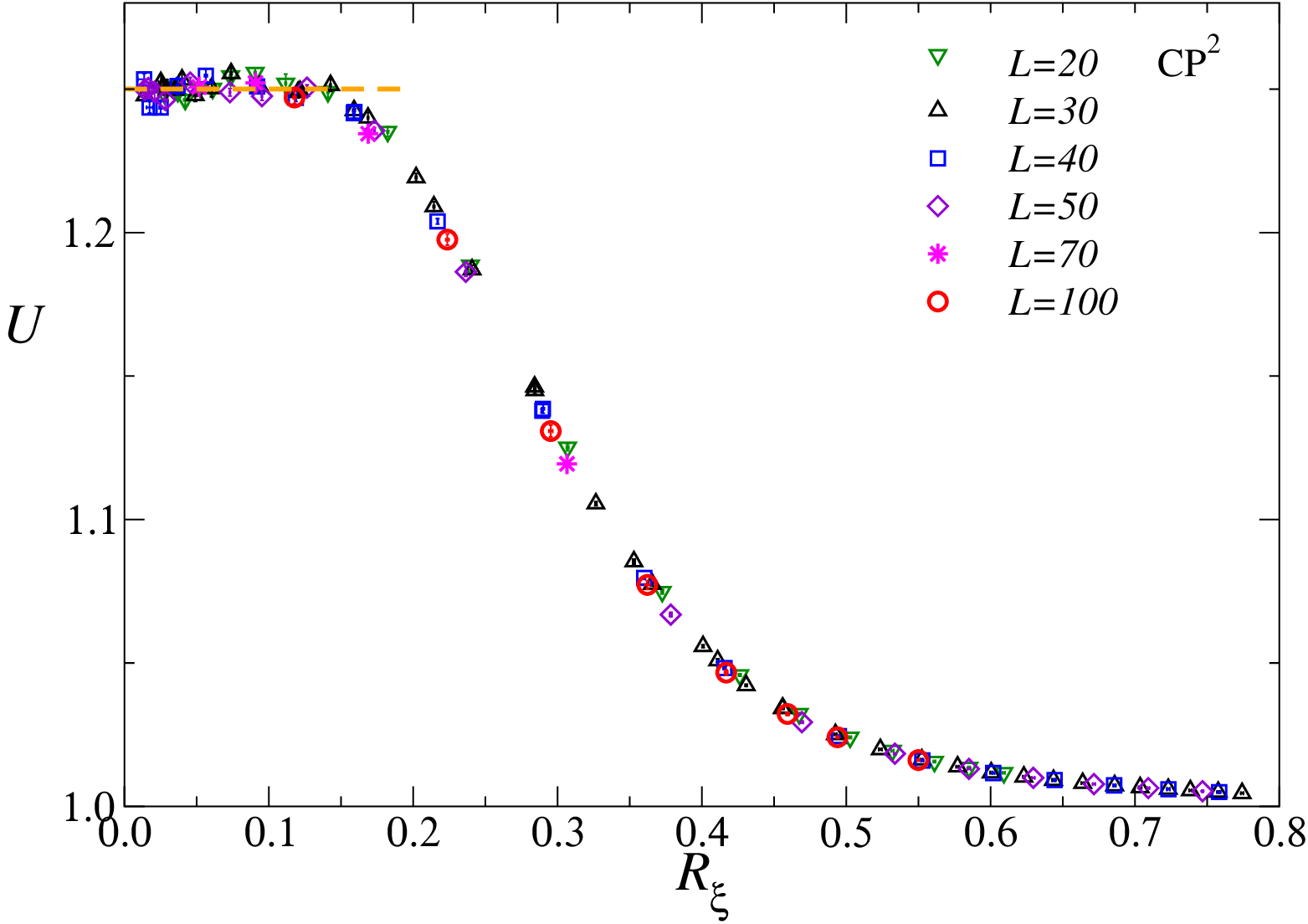}
\caption{Estimates of $U$ versus $R_\xi$ for the lattice CP$^2$ model
  obtained by setting $\gamma=0$ in the $N=3$ AH model (\ref{gllf}).
  The horizontal dashed line shows the asymptotic value 
  $U =5/4$ for $R_\xi\to 0$.  }
\label{bi-rxi-cp2}
\end{figure}

\begin{figure}[tbp]
\includegraphics*[scale=\graphicscale]{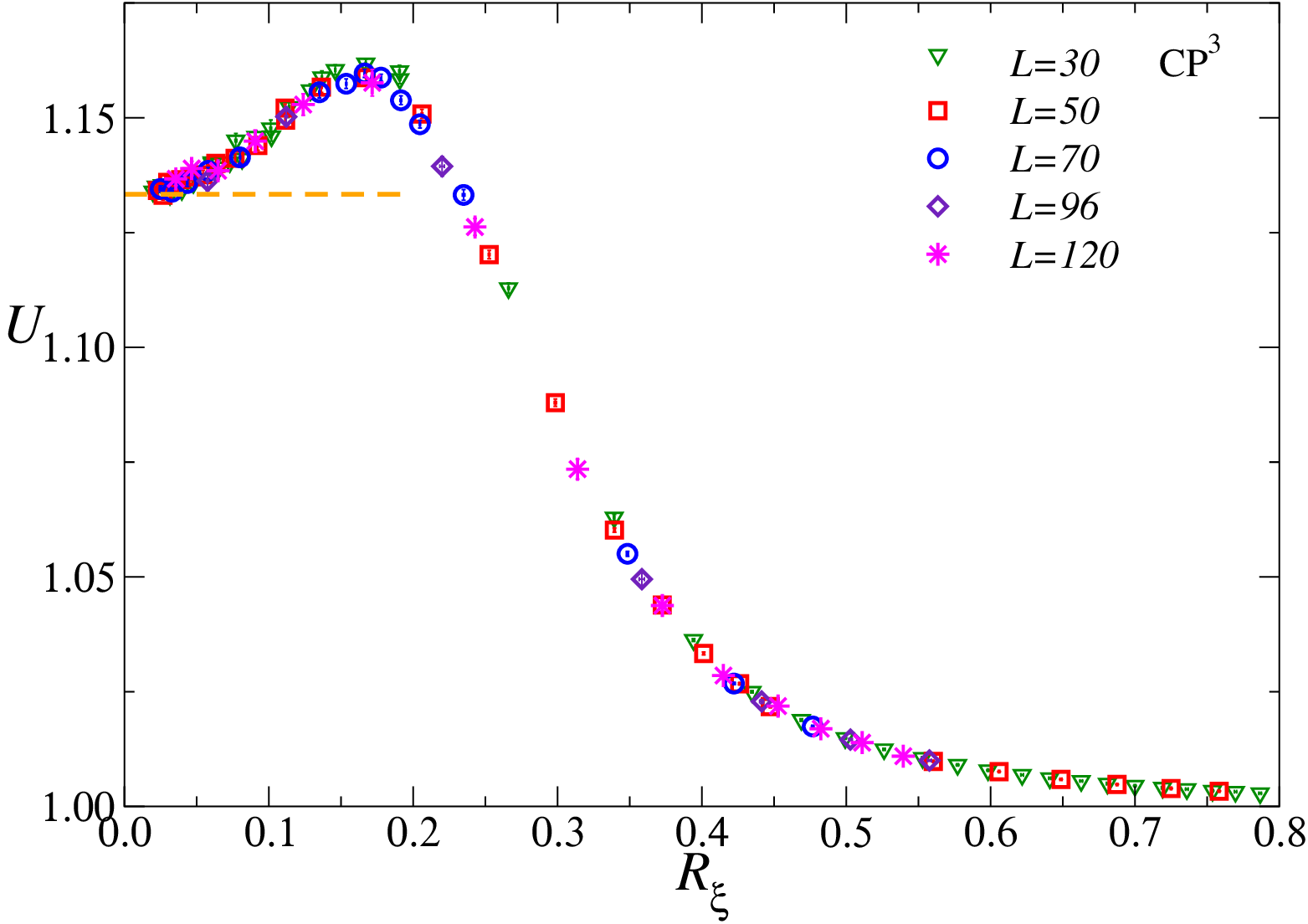}
\caption{Estimates of  $U$ versus $R_\xi$ for the lattice CP$^3$ model
($\gamma=0$).
  The horizontal dashed line corresponds the asymptotic value 
  $U=17/15$  for $R_\xi\to 0$.  }
\label{bi-rxi-cp3}
\end{figure}

To investigate the critical behavior of the model, we have performed
simulations for $N=2$ and $N=10$ and studied the FSS behavior of RG
invariant observables for different values of $\gamma$.  We find that
the asymptotic small-temperature behavior for generic finite values of
$\gamma$ is independent of $\gamma$ (at least for $\gamma$ not too
negative).  Therefore, the 2D AH model belongs to the universality
class of the 2D CP$^{N-1}$ model, obtained for $\gamma =
0$. Therefore, FSS curves and dimensionless RG invariant combinations
of observables in the thermodynamic and in the FSS limit should be
independent of $\gamma$.  Morover, $\xi$ should always increase as
$e^{c/T}$.  We expect this scenario to occur for any $N\ge 2$. For
instance, for any $\gamma$ the Binder parameter should have the same
FSS curve as in the lattice CP$^{N-1}$ models. Thus, for $N=3$ and
$N=4$, the universal behavior should be that shown in
Figs.~\ref{bi-rxi-cp2} and ~\ref{bi-rxi-cp3}, where we report results
for $\gamma=0$.

The independence of the critical behavior on $\gamma$ is related to
the subleading behavior of the gauge modes. For $\beta\to\infty$ also
vector and gauge correlations order, but in a much slower fashion. In
particular, in the infinite-volume limit $\xi_v/\xi \to 0$ as $\beta
\to \infty$, indicating that the critical behavior is completely
determined by the modes encoded in the gauge-invariant quantity
$Q_{\bm x}^{ab}$.

Summarizing, our numerical results show that the continuum 2D CP$^{N-1}$ field
theory~\cite{ZJ-book} 
provides the stable fixed point of the RG flow of the 2D lattice
AH model for any finite, not too negative $\gamma$.
Note that, in the limit $\gamma\to\infty$, the AH model becomes 
equivalent to the O($2N$) $\sigma$ model. Therefore, in this limit we expect
significant crossover phenomena, controlled by the unstable O($2N)$ fixed
point.

It is worth noting that the behavior of the 2D AH model 
presents some analogies with
that of its 3D counterpart. Indeed, also in three dimensions 
the nature of the
transition and the critical behavior, along the line separating the
two phases, is the same for any finite $\gamma$~\cite{PV-19-2}.
We mention that a
similar behavior also emerges in multiflavor scalar
chromodynamics, i.e., in lattice nonabelian gauge theories with
multicomponent scalar fields~\cite{BPV-19}. 
%However, unlike 3D lattice
%AH models where the gauge correlations give only rise to crossover
%effects along the transition lines, remaining noncritical, in the case
%of the 2D lattice AH models they also show asymptotic critical
%behaviors in the zero-temperature limit, determining the nature of the
%universal CP$^{N-1}$ critical behaviors.

As already mentioned in the introduction, 2D CP$^{N-1}$ models
present nontrivial topological properties similar to those of
QCD. Issues related to topology have been largely investigated, both
analytically and numerically, see,
e.g., Refs.~\cite{BP-85,CR-91,CRV-92,BBHN-96,v-99,ACDP-00,DMV-04,%%
  VP-09,Hasenbusch-17,BBD-19}.  The 2D multicomponent lattice AH model
is expected to share the same topological properties of CP$^{N-1}$
models.  In this paper we do not pursue these issues further, although
we believe that they are worth being investigated within the general
lattice AH model.


\begin{thebibliography}{99}
\bibitem{footnote-interpolation} An interpolation of the MC data of
  $U$ versus $R_\xi$ for the lattice 2D O(3) vector model is provided
  by the function 
$f(x)=1.66661 - (0.05343 \,x + 15.49447 \,x^2) (1 - e^{-10 x}) -
1.72086 \,x^2 + 62.76441 \, x^3 - 97.24352 \, x^4 + 71.99411 
\, x^5 - 20.96845 \, x^6$
%$f(x) = 1.66089 + 1.33868 \, x \, (1 - e^{-20\,x}) -
%  23.87085\,x^2 + 75.27449\,x^3 - 108.08162 x^4 + 75.65739\,x^5 -
%  21.01081\,x^6 $. It is
valid for $x\in [0,0.8]$. The error is
smaller than 0.5\%.
\end{thebibliography}
\end{document}